\documentclass[10pt]{article}

\usepackage{amsmath}
\usepackage{amssymb}

\usepackage{graphicx}

\usepackage{cite}

\usepackage{color}

\usepackage[labelfont=bf,labelsep=period,justification=raggedright]{caption}
\usepackage{indentfirst}

\makeatletter
\renewcommand{\@biblabel}[1]{\quad#1.}
\makeatother

\date{}

\begin{document}

% Title must be 150 characters or less
\begin{flushleft}
{\Large
\textbf{Two Categories of Indoor Interactive Dynamics of a Large-scale Human Population in a WiFi covered university campus}
}
% Insert Author names, affiliations and corresponding author email.
\\
Yi-Qing Zhang,
Xiang Li$^{\ast}$,
Lin Wang,
Yan Zhang\\

Adaptive Networks and Control Lab, Department of Electronic Engineering, Fudan University, Shanghai 200433,PR China

$\ast$ lix@fudan.edu.cn;
\end{flushleft}

% Please keep the abstract between 250 and 300 words
\section*{Abstract}

\textbf{Background}: The analyses of human daily interactive behaviors at the population level play a critical role to understand many dynamic processes, e.g., information diffusion, and disease transmission, taking place in the human society. The recent advance of information technologies such as mobile phone, RFID, wireless sensor, and our interest in, WiFi provides powerful measures to digitize the population interactions and facilitate quantitative investigations.\\

\textbf{Methods and Findings}: To explore large-scale population indoor interactions, we analyze \emph{18,715} users' WiFi access logs recorded in a Chinese university campus during \emph{3} months, and define two categories of human interactions, the event interaction (EI) and the temporal interaction (TI). The EI helps construct a transmission graph, and the TI helps build an interval graph. The dynamics of EIs show that their active durations are truncated power-law distributed, which is independent on the number of involved individuals. The transmission duration presents a truncated power-law behavior at the daily timescale with weekly periodicity. Besides, those \emph{`leaf'} individuals in the aggregated contact network may participate in the \emph{`super-connecting cliques'} in the aggregated transmission graph. Analyzing the dynamics of the interval graph, we find that the probability distribution of TIs' inter-event duration also displays a truncated power-law pattern at the daily timescale with weekly periodicity, while the pairwise individuals with burst interactions are prone to randomly select their interactive locations, and those individuals with periodic interactions have preferred interactive locations.\\

\textbf{Conclusion and Significance}: The dynamics of a large-scale population's indoor interactions shows the convolution of human activities' daily bursts and weekly rhythms. The `leaf' individuals in the aggregated contact network gathering to function as `hubs' in the transmission graph highlights the significant difference between aggregated and temporal structures, and the analyses of temporal dyads from the spatiotemporal perspective pave a new path to predict human interactive behaviors.

% Please keep the Author Summary between 150 and 200 words
% Use first person. PLoS ONE authors please skip this step.
% Author Summary not valid for PLoS ONE submissions.
%\section*{Author Summary}

\section*{Introduction}
Nowadays, a great deal of digital technologies are unobtrusively embedded into the physical world of human daily activities: we communicate with each other through emails and/or mobile phones, pay the shopping with credit cards, use public transportation by transit cards, and surf via wired or wireless networks. These technical innovations not only reshape our social life, but also record tremendous digital trajectories of human activities, which provide proxy clues to understand human behavioral patterns \cite{TechnoSocialSystem,ComputeSocialScience,email,mobilephone,banknote,onlinedata,Pentland2010AmerScience,Pentland2009WEF,Pentland2008Book,EagleN2005PhD,EagleN2006PUC,EagleN2009PNAS,RFID2010PLoSONE,RFID2011PLoSONEHosptial,RFID2010JTB,RFID2011PLoSONESchool,WirelessSensor2010PNAS,Takaguchi11PRX}. For instance, Pentland et al. \cite{Pentland2010AmerScience,Pentland2009WEF,Pentland2008Book} constructed `sociometers' to capture human honest signals in the workplace to infer their productivity and creativity. Eagle et al. \cite{EagleN2005PhD,EagleN2006PUC,EagleN2009PNAS} used Bluetooth embedded in the mobile phones to infer friendships from close proximity interactions. Barrat and his colleagues \cite{RFID2010PLoSONE,RFID2011PLoSONEHosptial,RFID2010JTB,RFID2011PLoSONESchool} built a flexible framework based on the distributed active Radio Frequency Identification (RFID) technology to record human face-to-face interactions in different rendezvouses, e.g., conference, office, hospital, museum, and school. Salath\'{e} et al. \cite{WirelessSensor2010PNAS} developed wireless sensors to evaluate the respiratory disease transmission risk among the members of a high school through person-to-person contacts. Takaguchi and his colleagues analyzed two sets of face-to-face interaction logs sampled by infrared sensor devices to predict conversation patterns\cite{Takaguchi11PRX}. Since human physical close proximity has been recognized to play a crucial role to (biological and computer) virus transmission, word-of-mouth information diffusion, and human social ties\cite{EagleN2009PNAS,RFID2011PLoSONEHosptial,RFID2010JTB,RFID2011PLoSONESchool,WirelessSensor2010PNAS,Takaguchi11PRX,RFIDDisease,RFIDInformation}, these fruitful researches not only shed light on human interactive features in a small community, but also inspire us to put a step further to explore human indoor interactions in an environment with a large-scale `natural' population.

Recent years, WiFi(or known as IEEE 802.11), as a ubiquitous wireless data exchange technology, has become one of the brightest areas in the communication business. Actually, WiFi signals can be found at almost every corner of the urban areas, and the notion of `WiFi-city' becomes reality. A university is a snapshot of the modern society, so is the WiFi coverage, where the WiFi control system records the digital access logs of the authorized WiFi users during their activities in the campus. Such WiFi access records work the proxy of a large-scale population's interactive activities with time stamps \cite{EPL68002,UrbComp2012}, which deserve more extensive efforts to capture the temporal patterns of their dynamics.

The WiFi dataset involved in this study was collected at Handan campus of Fudan University in Shanghai, China, which contains \emph{18,715} WiFi users' individual behavioral trajectories during \emph{3} months of a fall semester. We define two categories of human indoor interactions: event interaction(EI) and temporal interaction(TI). From the perspective of temporal networks\cite{Temporalnetworks}, the EIs help construct a transmission graph to characterize the group interactive features, and the TIs build an interval graph to characterize the dyad interactive features. The dynamics of EIs show that their active durations are truncated power-law distributed and `size-free', independent on the number of involved individuals. The transmission duration presents a power-law behavior at the daily timescale with weekly periodicity. Besides, some \emph{`leaf'} individuals in the aggregated contact network participate in the \emph{`super-connecting cliques'} of the aggregated transmission graph. Analyzing the dynamics of the interval graph, we find that the probability distribution of TIs' inter-event duration also displays a truncated power-law pattern at the daily timescale with weekly periodicity, while the pairwise individuals with burst interactions are prone to randomly select their interactive locations, and those individuals with periodic interactions have preferred interactive locations.

\section*{Results}

\subsection*{Two Categories of Human Indoor Interactions and Temporal Networks}

Before transforming the WiFi data of access logs as the proxy of a large-scale population's indoor interactions, we state our main assumption \cite{opportunistic2007IEEE,EPL68002,UrbComp2012} in this paper:  the WiFi devices seeing the same wireless access point(WAP) infers an indoor interaction among these devices' owners (see \emph{Methods} and \emph{Text SI} for the validity of the main assumption).

\begin{figure}[!ht]
\begin{center}
\includegraphics[width=4in]{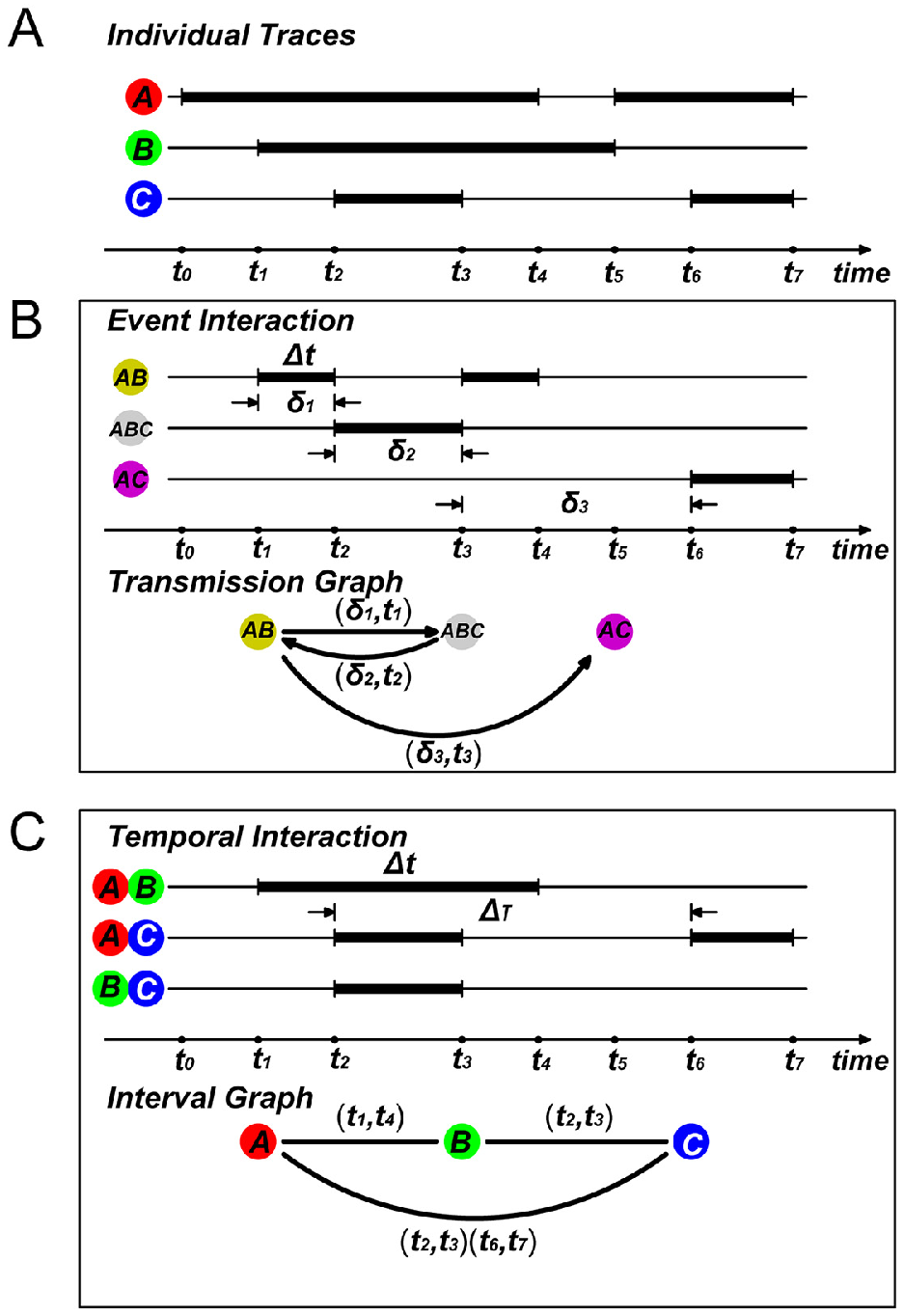}
\end{center}
\caption{
{\bf Two Categories of Human Indoor Interaction and Temporal Networks.} (a)WiFi access logs are translated into individual behavioral trajectories. The bold line pertains to their online duration. (b) Event interaction is defined to catch the group interactive dynamics and evolution features.  Each bold line possesses an exclusive time interval, where the corresponding individuals are assembled into a contact clique. The transmission graph gives a reduced picture of the event interactions, where the nodes pertain to the event interactions (contact cliques), and the edges illustrate the transmission paths between different EIs. (c)Temporal interaction is defined to describe the dyad contacts. Each temporal interaction(bold line) is a person-to-person encounter. The interval graph gives a reduced picture of temporal interactions, where vertices are individuals, and each edge represents at least one TI.
}
\label{schematic}
\end{figure}

We first define two categories of human indoor interactions and temporal graphs. Figure \ref{schematic}(a) shows an instance which translates three WiFi users' access logs into the individual behavioral traces. The bold lines pertain to their online durations. In the observed epoch [$t_0,t_7$], these three users are uniquely recorded by this WAP.

\indent

\textbf{From Event Interactions To Transmission Graph}

\indent

With the aforementioned main assumption - the users 'seeing' the same WAP have an indoor interaction, Figure \ref{schematic}(b) illustrate the process of translating individual behavioral trajectories into event interactions(EIs) and constructing the corresponding transmission graph(TG). Event interaction is defined to characterize the group interactive dynamics and evolution features, which facilitates the analysis of epidemic spread or information diffusion. For instance, in the epoch of [$t_{1},t_{2}$], the online duration of \emph{A} and \emph{B} are overlapped, we therefore define an EI $E_{AB}^{t_1}$ characterizing the fact that \emph{A} and \emph{B} are assembled as a contact group.

Regarding event interactions as vertices \cite{TG1,TG2}, we define two vertices are linked given the following three rules are satisfied:

\begin{enumerate}
  \item In the time series, a source EI (e.g., $E_{AB}^{t_1}$) is the closest EI prior to the sink EI ($E_{ABC}^{t_2}$).
  \item At least one user coexists in the source and sink EIs.
  \item When there are several sources before one sink, any set of the shared users between the given source and sink EIs never intersect with each other(set).
\end{enumerate}

The three rules ensure that the source EI has a potential to transfer some information or virus to the sink EI due to the intermediate role of shared individuals. As illustrated in Figure 1(b), $E_{AB}^{t_1}$ is active in the epoch [$t_{1},t_{2}$] and $E_{ABC}^{t_2}$ in [$t_{2},t_{3}$], while two EIs satisfy the above three rules, a directed edge links these two vertices in the transmission graph.

To quantify the temporal information in event interactions and the transmission graph, we introduce the following quantities:

\begin{itemize}
  \item The size \emph{s} of a given EI is defined as the number of users involved.
  \item If a given EI is active in the period of [$t^{begin},t^{end}$], its active duration is defined as \\$\Delta t^{EI}=t^{end}-t^{begin}$.
  \item Given two connected EIs, $i, j$, linked in the TG, denote the edge weight $\omega_{ij}=(\delta, t^{observed})$, i.e., the shortest transmission duration from the source $E_i$ to the sink $E_j$ is $\delta$ when observed at the time $t^{observed}$. (Without special statement, $t^{observed}$ is considered as $t^{begin}$ of each source EI in this paper.)
\end{itemize}

Generally, the topological properties of a temporal graph are measured from the corresponding aggregated network on a given time window \cite{RFID2010PLoSONE,RFID2011PLoSONEHosptial,RFID2010JTB,RFID2011PLoSONESchool }. Here, we aggregate the transmission graphs on the whole observation period (\emph{3 } months) to generate the aggregated transmission graph (ATG), and define several quantities as follows:

\begin{itemize}
  \item The frequency $n^{EI}$ gives the number of EI's active times. For instance, $E_{AB}$ in Figure \ref{schematic}(b) takes place in two time intervals, $[t_1,t_2]$ and $[t_3,t_4]$, so the frequency $n^{EI}_{AB}=2$.
  \item The number of repeated transmission paths from source EI (e.g.,$E_i$) to sink EI ($E_j$) is frequency $n^{TP}_{E_iE_j}$.
\end{itemize}

\indent

\textbf{From Temporal Interactions To Interval Graph}

\indent

We further define temporal interactions to characterize the dyad interactions. With the main assumption mentioned above, \emph{A} and \emph{B} in Figure \ref{schematic}(a,c), who share a common duration, assemble into a temporal interaction $T_{AB}^{t_1}$ during the epoch [$t_{1},t_{4}$]. The set of temporal interactions(TIs) can be represented by an interval graph \cite{Temporalnetworks} directly, where users are the vertices, and TIs are the edges.

The following definitions of an interval graph are introduced to quantify the temporal information of dyad interactions:

\begin{itemize}
  \item Given an active TI during [$t^{started},t^{finished}$], the active duration of the edge (TI) is \\$\Delta t^{TI}= t^{started}-t^{finished}$.
  \item The frequency of edge $n^{TI}_{ij}$ gives the number of times that two connected users $i,j$ repeat their TIs.
  \item Given the same dyad's two successive TIs, $T^{1}_{ij}$ and $T^{2}_{ij}$ ($i,j$ are two users), their active durations are [$t^{started}_{1},t^{end}_{1}$] and [$t^{started}_{2},t^{end}_{2}$], respectively. The inter-event duration is defined as the difference between the beginning time of these two successive TIs ($\Delta\tau=t^{started}_{2}-t^{started}_{1}$). Take \emph{A} and \emph{C} in Figure \ref{schematic}(c) as an example, $\Delta\tau_{AC}=t_{6}-t_{2}$.
\end{itemize}

The dyads' TIs have a non-decreasing time sequence if their frequency $n^{TI}_{ij}>1$. As shown in Figure \ref{schematic}(c), the TIs of \emph{A,C} have a non-decreasing active time sequence:[$t_2,t_3$],[$t_6,t_7$].

\indent

To summarize, as shown in Figure \ref{schematic}, event interactions are driven by the group access-related events, while temporal interactions are driven by the time sequences of dyad interactions. Each event interaction possesses an exclusive time interval on the WAP it takes place, whereas different temporal interactions may coexist with each other (e.g., $T_{AC}^{t_2}$ and $T_{BC}^{t_3}$ in Figure \ref{schematic}(c)).

\subsection*{Dynamics of event interaction}

\begin{figure}[!ht]
\begin{center}
\includegraphics[width=4in]{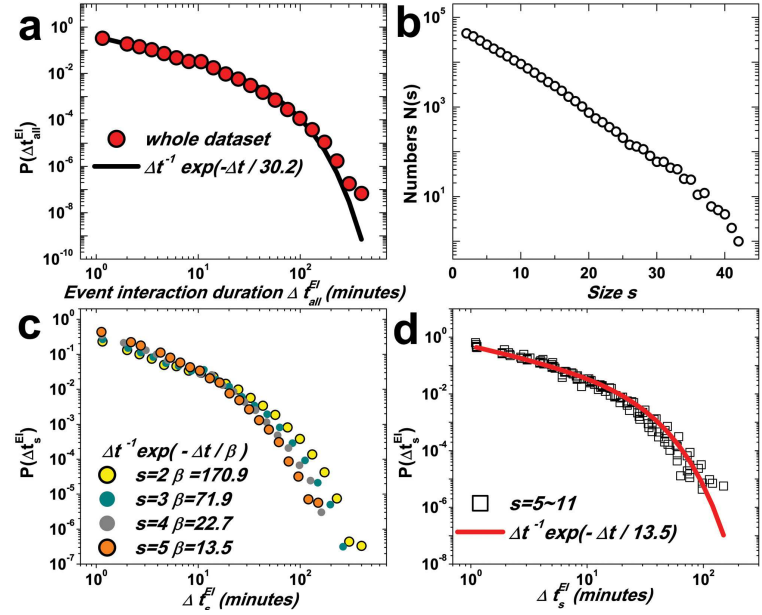}
\end{center}
\caption{
{\bf Dynamics of the Event Interactions(EIs).} (a) The probability distribution of all the EIs' active duration $\Delta t^{EI}_{all}$. (b) The distribution of the size of different event interactions. (c) The probability distribution of the EIs' active durations $\Delta t^{EI}_{s}$ with a given size $s = 2,3,4,5$. (d) The probability distribution of the EIs' active durations $\Delta t^{EI}_{s}$ with a given size $s=5,6,...11$.
}
\label{dynei}
\end{figure}

Figure \ref{dynei}(a) shows that the probability distribution of EIs' active durations collected in 3 months is a truncated power law (the exponent of the power-law part approximates to 1), indicating that the long-lasting group interactions can hardly survive due to the frequent emergence of abundant short-active-duration events. In Figure \ref{dynei}(b), the number of event interactions with a given size (\emph{N(s)}) is exponential distributed, which further implies that the EIs having large sizes are rare. Therefore, it is natural to explore the dependence between the distribution of EIs' active durations $\Delta t^{EI}$ and a given size \emph{s}. Firstly, we study the impact of small sizes on EIs' active durations: the probability distributions of EIs' active durations $\Delta t^{EI}_{s}$ with size $s=2,3,4,5$ exhibit the truncated power-law behaviors, and their power-law parts present the similar exponents close to 1. However, their exponential cutoffs gradually decay with the growth of event size(Figure \ref{dynei}(c)). Besides, when the size of EIs is larger than 5, the influence of EIs' sizes to their active durations is subtle. Figure \ref{dynei}(d) reports the probability distributions of the EIs' active durations with a given size $s\geq 5$, which keeps almost invariant with the growth of size, i.e., the active durations of EIs are \emph{`size free'} in this case.

\begin{figure}[!ht]
\begin{center}
\includegraphics[width=4in]{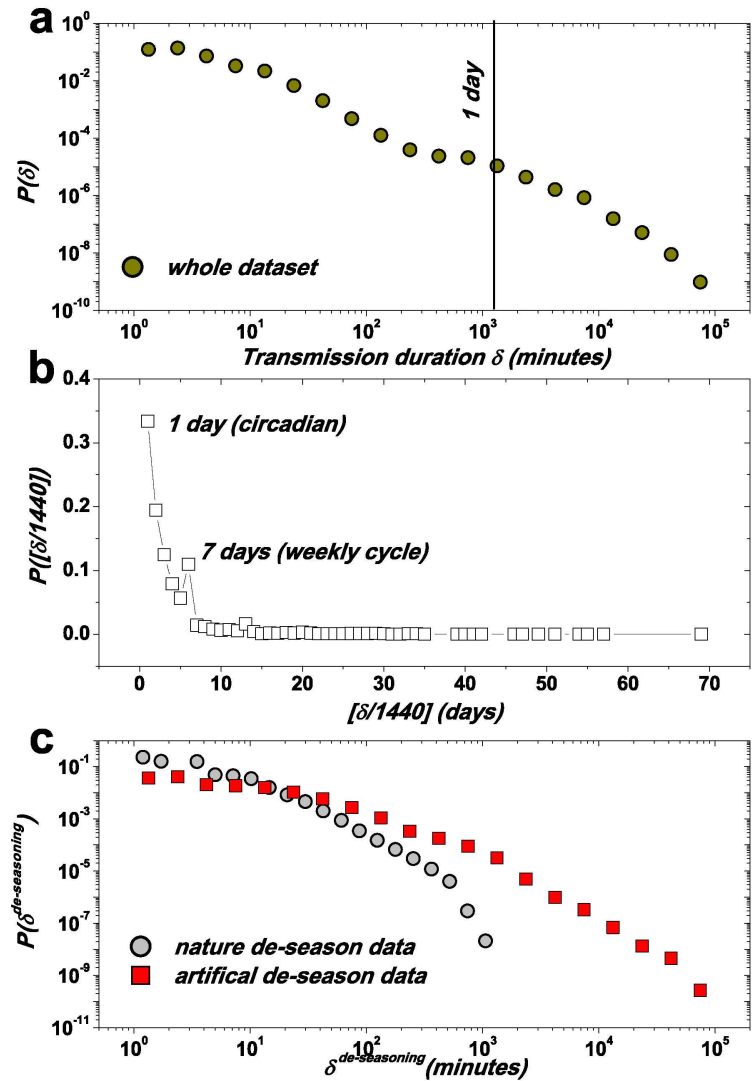}
\end{center}
\caption{
{\bf Dynamics of the Transmission Graph.} (a) The probability distribution of all transmission durations $\delta$.(b) The probability distribution of integral days spent by transmission duration $[\frac{\delta}{1440}]$ (days).(c) The probability distribution of transmission durations under the natural de-seasoning method (gray circle) and the artificial de-seasoning time-shuffled method (red square).
}
\label{dyntg}
\end{figure}

As shown in Figure \ref{dynei}, all EIs' active durations are less than 5 hours, which are much less than the circadian rhythms \cite{Circadian2007}. On the contrary, many transmission durations are obviously longer than 24 hours. In Figure \ref{dyntg}(a), the probability distribution of the transmission durations in the whole dataset falls into a bifold power-law with the turning point approximately equal to 24 hours, indicating the existence of circadian rhythms. We also calculate the integral days spent by each $\delta$ ([$\frac{\delta}{1440}$]) to examine whether some periodicity exists. In Figure \ref{dyntg}(b), the probability distribution of [$\frac{\delta}{1440}$] has two peaks at the 1st day and 7th day, i.e., the daily bursty behavior of successive event interactions coevolutes with weekly rhythms. We further perform the de-seasoning analysis to remove the circadian or weekly rhythms presented in Figure \ref{dyntg}(a,b) with two methods: I) natural de-seasoning method: only conserve the source and sink EIs taking place in the same day; II) artificial de-seasoning time-shuffled method in \cite{Temporalnetworks,Karsai11PRE,Jo12NJP,KivelaArxiv2012} to suit event interactions' definition (the details see \emph{Text SI}). As shown in Figure \ref{dyntg}(c), the probability distributions of the transmission durations 'filtered' by two de-seasoning methods both fall into the truncated power-laws, which indicates that the bursty behavior of event interactions takes place independently on the daily and weekly rhythms.

In addition, we examine the dependence of the source and sink EIs' active durations of each transmission path as shown in Table \ref{dependence}. Both the Pearson's coefficients and memory coefficients \cite{burstcoefficient} between them are very minor, indicating there does not exist the interdependence although the two groups of EIs share some individuals. From the definition of transmission graph, the transmission durations may be equal to the source EIs' active durations. However, their Pearson's coefficients and memory coefficients are also very small, hence the source EI' active duration is only the lower bound of the corresponding transmission duration, and there is no more dependence between them.

\begin{figure}[!ht]
\begin{center}
\includegraphics[width=4in]{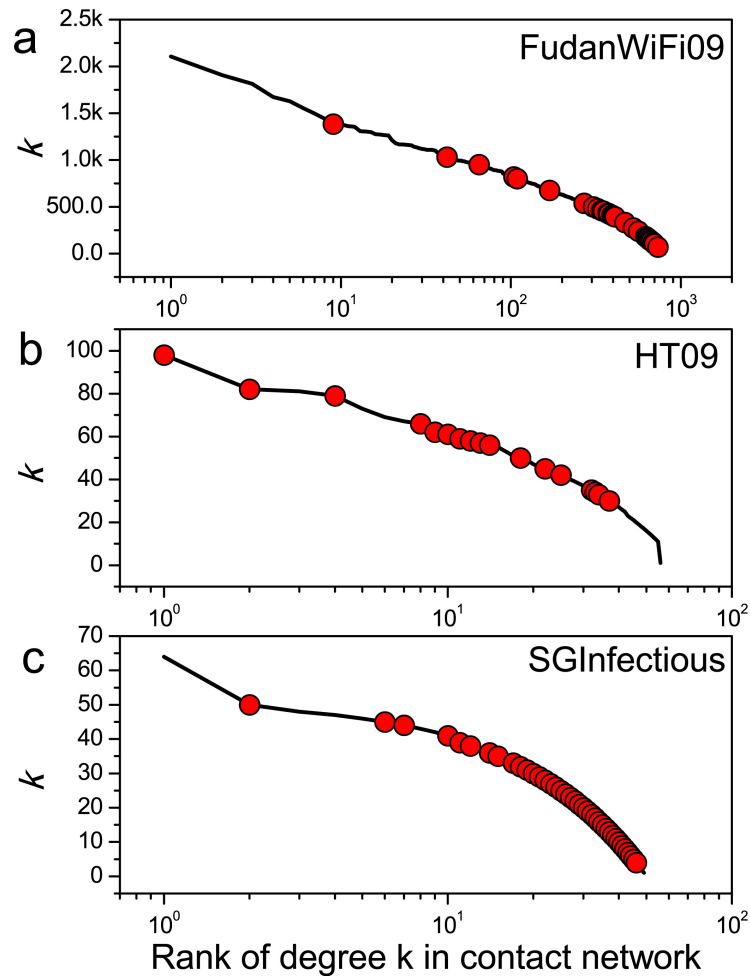}
\end{center}
\caption{
{\bf The Degree of Super-connecting Groups' Members Versus Their Ranks in the Aggregated Contact Network.} (a) The degrees versus their ranks in the traditional aggregated contact network(the vertices exhibits people) produced by `FudanWiFi09'. The members of super-connecting groups are shown by the red circle. (b,c) The degrees versus their ranks $r$ in the traditional aggregated contact network produced by `HT09' and `SGInfectious', respectively.
}
\label{leafhub}
\end{figure}

We construct the aggregated transmission graph (ATG) with the whole 3-month WiFi data set (the statistical properties of the ATG see Figure S1 and \emph{Text SI}). In Figure S1(a), most of the vertices only have small out-degrees ($k_{out}<10$) and in-degrees($k_{in}<10$), but there are 20 vertices with both high out-degrees and in-degrees to function as `super-connecting groups'(hubs) (see \emph{Text SI}). Recent studies reveal that super-connectors(hubs) in an aggregated network dominate the susceptible-infected-susceptible spreading processes due to the fact that they are self-sustained sources continually diffusing the virus to other neighbors \cite{SpreadingPRL2001,ABRMP,NewmanSIAM,PhysRep,AdvanInPhys,NewmanPRE2002,ThresholdPRL2010}. Therefore, the members of these `super-connecting groups' play a crucial role in the dynamics of disease transmission. However, traditional studies about the roles of individuals in the spreading processes focus on human contact network (for reviews see \cite{spreadingReview1,spreadingReview2}), where vertices are individuals and links are their contacts. We also aggregate the whole WiFi data set into a contact network(the detailed aggregation procedure refers to \emph{Text SI}) and find that the members of those `super-connecting groups' are `leaf' individuals in the contact network(red circle in Figure \ref{leafhub}(a)). Moreover, we further explore two other human indoor interaction data sets collected in a conference and an public exhibition\cite{RFID2010JTB}(see \emph{Methods and Materials}). Both these two high-resolution human face-to-face data sets(Figure \ref{leafhub}(b,c)) and our data set (Figure \ref{leafhub}(a)) illustrate that `leaf' individuals in the static human contact network gather to function as `hubs' in the transmission graph independently on different rendezvouses, indicating that the roles of many `leaf' individuals in a traditional static contact network may have been underestimated in a spreading process.

\subsection*{Dynamics of temporal interaction}
\begin{figure}[!ht]
\begin{center}
\includegraphics[width=4in]{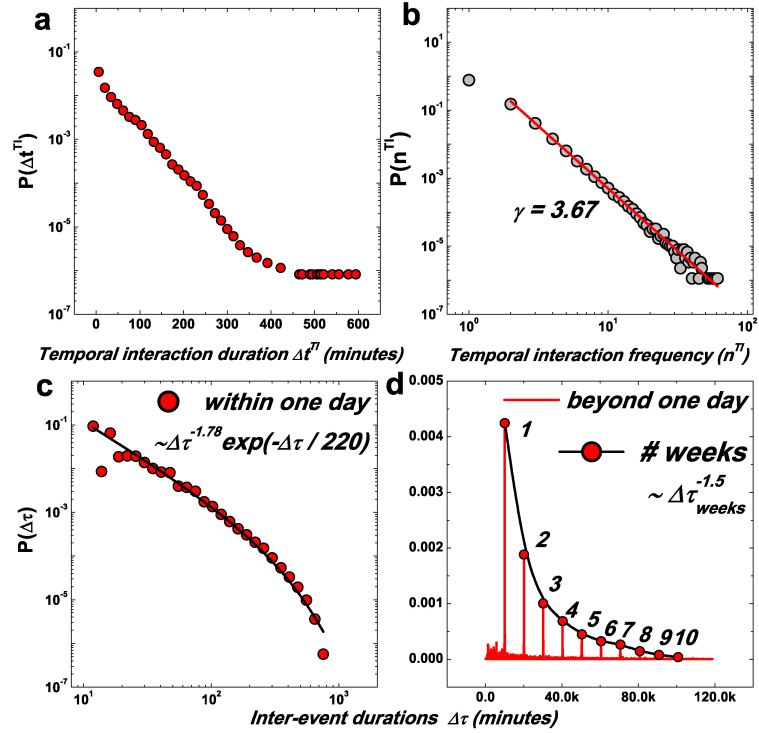}
\end{center}
\caption{
{\bf Dynamics of the Temporal Interactions(TIs).} (a) The probability distribution of TIs' active duration $\Delta t^{TI}$. (b) The probability distribution of TIs' frequency $n^{TI}$. (c) The probability distribution of inter-event durations $\Delta \tau$ of the successive TIs that are active in the same day. (d) The probability distribution of inter-event durations $\Delta \tau$ of the successive TIs that are active beyond one day. The probability distribution of its peaks exhibits a power-law feature.
}
\label{dynti}
\end{figure}

As shown in Figure \ref{dynti}(a), the probability distribution of TIs' active durations shows an exponential pattern(and the maximum value of $\Delta t^{TI}$ is less than 24 hours), which is different from the empirical results of human close proximity interactions recorded by Bluetooth \cite{EagleN2006PUC} or wireless sensors \cite{WirelessSensor2010PNAS}. This difference might derive from the fact that our data reflects the university daily teaching schedules and the human circadian rhythms, which constrain the individuals' resident periods in classrooms. Since the TIs' active durations can not reflect the population's interactive features beyond one day, we analyze the probability distribution $P(n^{TI})$ of TIs' frequencies in Figure \ref{dynti}(b). The outcome of power-law distribution indicates that although the number of the frequent contacting pairwise individuals is far less than that of the casual encounters, they can not be ignored.

The inter-event duration measures the time interval between two successive TIs of the same dyads (considering the start point of the two TIs). We also partition all the inter-event durations into two subsets according to the following criterion: whether or not two TIs of the same two users take place in the same day. As shown in Figure \ref{dynti}(c), the probability distribution of the inter-event durations of the successive TIs taking place in the same day exhibits a truncated power-law, which is similar to the findings in \cite{interContact2007,opportunistic2007IEEE,opportunistic2005,pocketSwitched2005,Eagle2007}. Moreover, Figure \ref{dynti}(d) reports the probability distribution of the inter-event duration of the successive TIs taking place in different days, where the red circles emphasize that there are many remarkably peaks indicating the weekly periodicity of the inter-event features. Therefore, the temporal interactions is in fact the convolution of burst behaviors on daily timescale and weekly rhythms.

\begin{figure}[!ht]
\begin{center}
\includegraphics[width=4in]{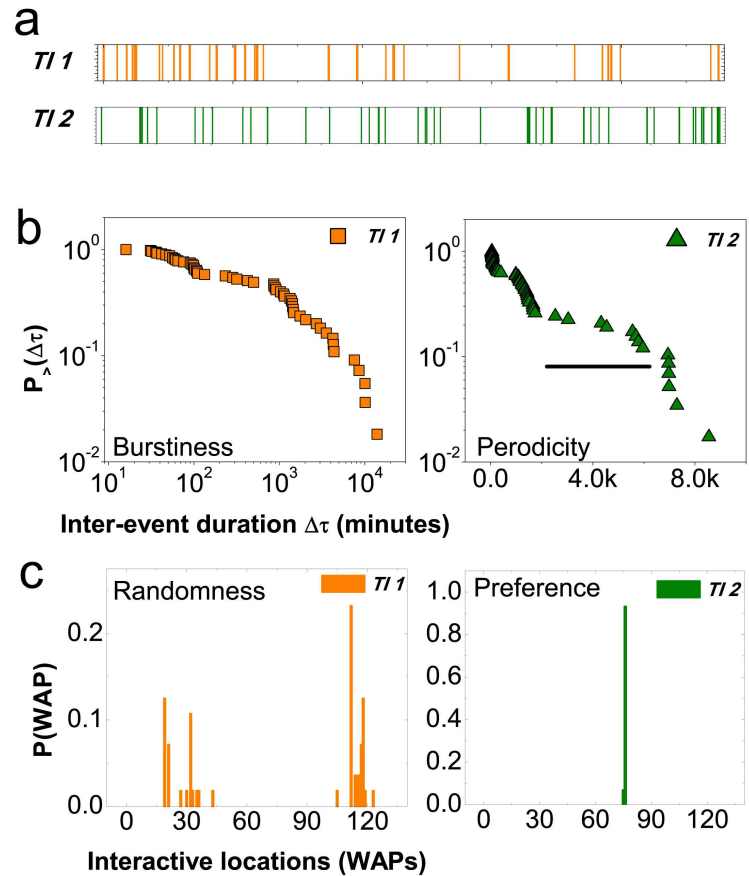}
\end{center}
\caption{
{\bf Spatiotemporal Correspondence in Temporal Interactions(TIs).} (a) The series of two typical TIs \emph{TI 1}(up) and \emph{TI 2}(bottom). The horizontal axis denotes time, each vertical line corresponds to an interaction. (b) The cumulative probability distribution of inter-event times  $\Delta \tau$ of \emph{TI 1}(orange square) and \emph{TI 2}(green triangle). (c) The occurrence probability of \emph{TI 1}(orange) and \emph{TI 2}(green) in corresponding interactive locations (WAPs/Voronoi cells). The \emph{`spatial entropy'} of \emph{TI 1} is 2.48 and the \emph{`spatial entropy'} of \emph{TI 2} is 0.25.
}
\label{spatiotemporal}
\end{figure}

To further explore the pairwise individuals' inter-event durations, we randomly choose two frequently interactive dyads. The two chosen series of temporal interactions are labeled as \emph{TI 1} and \emph{TI 2} with frequency $n^{TI}_1=56,n^{TI}_2=59$, respectively. In Figure  \ref{spatiotemporal}(a), the time series \emph{TI 1} shows a typical burst characteristics, while the time series \emph{TI 2} presents more regular intervals between two active events. We analyze the cumulative probability distribution(CPD) of the inter-event durations of \emph{TI 1, TI 2} in Figure \ref{spatiotemporal}(b). The CPD of \emph{TI 1}'s inter-event durations falls into a power-law with the cutoff of one week (10,000 minutes), while the CPD of \emph{TI 2}'s inter-event durations presents an exponential form. In addition, there does not exist the dependence between TI's active durations and inter-event durations as shown in Figure S3, which are uncorrelated.

Since the addresses of WAPs are encoded in the temporal interactions of pairwise users, we define '\emph{spatial entropy}' to measure the spatial information of involved temporal interactions:

$$H_{ij}(n^{TI})=-\sum\limits_{\ell=1}\limits^{L}p_{\ell} \ln p_{\ell}$$ ($L \leq n^{TI}$ and $n^{TI} \geq 2$), where $p_{\ell}$ is the probability that the TIs of users $i,j$ simultaneously occur at the location '$\ell$' (equivalent to WAP '$\ell$'). The average spatial entropy with a given frequency is averaged as:

$$<H(n^{TI})>=\frac{\sum\limits_{ij\in \Psi } H_{ij}(n^{TI})}{N(n^{TI})}$$ where $\Psi$ represents the set of pairwise users with the same frequency ($n^{TI}$), and $N(n^{TI})$ is the size of $\Psi$.

Figure S4 reports the average spatial entropy $<H(n^{TI})>$  increases with the growth of frequency $n^{TI}$ at the population level. However, at the level of dyads, the TIs' spatial entropy are independent on their frequency $n^{TI}$, as illustrated by two aforementioned (\emph{TI 1, TI 2}) in Figure \ref{spatiotemporal}(a). We further observe the occurrence probability of \emph{TI 1, TI 2} in Figure \ref{spatiotemporal}(c). The low entropy of \emph{TI 2} ($H = 0.25$) indicates that two individuals have preferred interactive locations, and the predication of their future interactive location is possible. While the high entropy of \emph{TI 1}($H = 2.48$) means that the involved individuals randomly select their interactive locations. To summarize, we conclude that the pair of individuals with burst interactions (\emph{TI 1} in Figure \ref{spatiotemporal}(b)) randomly select their interactive locations (Figure \ref{spatiotemporal}(c)), and the pair of individuals with periodic interactions (\emph{TI 2}) present their preference on the interactive locations.

\section*{Materials and Methods}

\subsection*{Ethics Statement}

Although there is no official institutional review board (IRB) or ethics committee in Fudan University, this research achieves the approvement from the Informatization Office of Fudan University, which is the chief office responsible for the deployment of the WiFi wireless access point network and data collection at Fudan University. According to the local regulation rules of Fudan University `THE REGULATIONS FOR ADMINISTRATION OF FUDAN UNIVERSITY WIFI CAMPUS USERS' (available: http://www.xxb.fudan.edu.cn/s/52/t/104/42/e6/info17126.htm (Chinese Edition)), the part of the ninth provision, `Without licence, monitor other people's information with any monitor software is forbidden.', we have signed a confidentiality agreement and obtained the permission from the Informatization Office of Fudan University, the administrative unit of department level in Chinese Government. We declare that we have the responsibility to protect the privacy of the WiFi Campus users at Fudan University involved in this research, the data set involved is de-identified, and none privacy information of the WiFi users are available.

\subsection*{Study Setting}

The WiFi wireless access point(WAP) network involved in this study is deployed at Handan campus of Fudan University in Shanghai, China. All of the WAPs deployed in the Campus provide the dual-band(802.11g/n (2.4-GHz) and 802.11a (5-GHz)) free wireless accessing services to all the authorized campus members (e.g., students, teachers, office staffs and visiting scholars). These members' wireless electronic devices are generally automatically connected to the \emph{closest} WAP. Only when the WAP is overloaded, the device will be switched to another adjacent WAP. In the campus WiFi system of Fudan University, each WAP can serve around 50 users. Considering the percentage of students equipping wireless devices in the involved data set collected in the teaching buildings, it is sufficient to guarantee that generally every wireless device is served by the closest WAP. Furthermore, the number of devices co-served by one WAP is equal to the size of event interactions. As shown in Figure \ref{dynei}(b), the maximum number of co-served devices is less than 50, i.e., all the involved WAPs were not overloaded in the three months.

\subsection*{The Pre-processing of the Data Set}

The main data set used in our study is named as `FudanWiFi09' which records the users' access logs in the 2009-2010 fall semester (18/10/2009-9/1/2010). It contains the hardware information of users' electronic devices (the Media Access Control address, CPU trademark, et al.), the hardware information of the connected WAP, and the connecting/disconnecting time as well. Since the Media Access Control address, as the unique serial number, can uniquely identify different electronic devices and WAPs, we conserve them with the connecting/disconnecting time in `FudanWiFi09'.

From the statistics of Fudan University (http://www.fudan.edu.cn/new\_genview/genview.htm), there are totally 33,239 students, teachers and staffs in the campus, and `FudanWiFi09' contains the records of 22,050 individuals' 423,422 behavioral trajectories in the three-month period. Since we target the proxy data of those wireless devices to feature the users' indoor interactive behaviors, we only focus our research on the WiFi data recorded in those buildings which are open to public without safeguards for the WiFi users' devices (i.e., when a user leaves such a place, he/she should bring the device with himself/herself.)  Therefore, the dataset of 18,715 individuals' 262,109  behavioral trajectories from all six teaching buildings with the WiFi coverage have been employed in the work of this paper.

\subsection*{Two Public Data Sets Used in the Study}

Two public data sets used in the study are both collected by the RFID technology available at the website of SocioPatterns (http://www.sociopatterns.org), and a related research refers to \cite{RFID2010JTB}. The first data set `HT09' (http://www.sociopatterns.org/datasets/infectious-sociopatterns-dynamic-contact-networks/) was collected during the ACM Hypertext 2009 conference, hosted by the Institute for Scientific Inter-change Foundation in Turin, Italy, from June 29th to July 1st. It involved about 100 conference participants to wear radio badges which monitored around 10,000 face-to-face interactions in the conference period of three days. The next data set `HGInfectious' (http://www.sociopatterns.org/datasets/hypertext-2009-dynamic-contact-network/) contains more than 230,000 face-to-face contacts among more than 14,000 visitors, which was collected during the art-science exhibition `INFECTIOUS: STAY AWAY' at the Science Gallery of Dublin, Ireland, from April 17th to July 17th, 2009.

These two data sets both have the similar tab-separated lists representing the active contacts during 20-second intervals of the data collection.  Each line has the form `$t$ $i$ $j$', where $i$ and $j$ are the anonymous IDs of the persons in contact, and the interval, during which this contact was active, is $[t-20s, t]$. In the study, the two data sets are pre-processed by assembling the continues contacts of the same pair into one interaction, before translating them into the defined event interactions as shown in Figure \ref{schematic}.

\subsection*{The Average Spatial Distances of Two Users}

In the campus WiFi system of Fudan University, each WAP can serve around 50 users, which we conjecture it is sufficient to guarantee that every device is served by the closest WAP. Besides, the WAPs in adjacent floors have none overlapped covered regions because the building materials can dramatically attenuate the wireless signals, and each floor can be decomposed by the WAPs into the corresponding Voronoi tessellations. However, it is impossible to precisely locate the position of a user within a Voronoi \cite{Gonzalez2008Nature}, and the spatial distance between any two users inside one Voronoi cell can not be directly calculated. Therefore, we propose an  approximative method to estimate the average spatial distance of any two users in Voronoi cells (The details see \emph{Text SI}), and Table \ref{distance} shows the average spatial distances of any two users and their standard derivation in all involved teaching buildings.

% Do NOT remove this, even if you are not including acknowledgments
\section*{Acknowledgments}
We gratefully acknowledge the insightful comments from Prof. Alain Barrat and two anonymous reviewers, and the assistance from Bochun Wu, and the Informatization Office of Fudan University for the WiFi Data collection, and the Archives of Fudan University for providing the blueprint of all the teaching buildings. This work was partly supported by the National Key Basic Research and Development Program (No.2010CB731403), the NCET program (No.NCET-09-0317), and the National Natural Science Foundation of China.

%\section*{References}

% The bibtex filename
%\bibliography{template}

\clearpage

\section*{Figure Legends}

\clearpage

\section*{Tables}

\begin{table}[!ht]
\caption{
\bf{The dependence of the source, sink EIs' active durations and transmission durations in the transmission graph}}
\begin{tabular}{|c|c|c|}
\hline
& Pearson's Coefficient & Memory coefficient\\\hline
The active durations of source EIs and sink EIs & 0.0497 & 0.1950 \\\hline
The active duration of source EIs and transmission duration & 0.0497 &0.0497\\\hline
\end{tabular}
\begin{flushleft}
\end{flushleft}
\label{dependence}
\end{table}

\begin{table}[!ht]
\caption{
\bf{Average spatial distances between any two users seeing the same wireless access point in each building.}}
\begin{tabular}{|c|c|c|c|}
\hline
Teaching Building & Number of WAPs & $<d>$(m) & $\delta$(m)\\\hline
the $2^{nd}$ bldg. & 16 & 6.26 & 1.96\\\hline
the $3^{rd}$ bldg. & 21 & 7.14 & 0.96\\\hline
the $4^{th}$ bldg. & 15 & 7.86 & 1.71\\\hline
the $5^{th}$ bldg. & 28 & 6.98 & 1.37\\\hline
the $6^{th}$ bldg. & 33 & 7.11 & 1.59\\\hline
the west bldg. & 14 & 11.75 & 0.98\\\hline
\end{tabular}
\begin{flushleft}$<d>$ is the average spatial distances and $\delta$ is the standard variation of spatial distances.
\end{flushleft}
\label{distance}
\end{table}

\end{document}